\documentclass[proceedings]{JHEP3} 
\PrHEP{ hep2001}

\usepackage{epsfig,multicol}			

\newbox\mybox
\newcommand\fverb{\setbox\mybox=\hbox\bgroup\verb}
\newcommand\fverbdo{\egroup\medskip\noindent\fbox{\unhbox\mybox}\ }
\newcommand\fverbit{\egroup\item[\fbox{\unhbox\mybox}]}

\title{Hadronic structure of the photon at LEP}

\author{{\'Akos Csilling}\thanks{On leave of absence from 
KFKI Research Institute for Particle and Nuclear Physics,
H-1525 Budapest, P.O.Box 49, Hungary}\\
	CERN, CH-1211, Geneva, Switzerland\\
	E-mail: \email{Akos.Csilling@cern.ch}}

\conference{International Europhysics Conference on High
                      Energy Physics}

\abstract{
Recent improvements in the analysis techniques and Monte Carlo models used
in the measurement of the hadronic structure function 
of the photon have lead to much improved
experimental results. Its low $x$ behaviour 
was studied in various $Q^2$ regions using LEP1 and LEP2 data,
while  its $x$ dependence and $Q^2$ evolution up to very high virtualities, 
as well as its charm component were studied using
the high energy and luminosity data of LEP2. 
These recent results will be presented.
}

\begin{document} 

\section{Introduction}

The photon is a unique particle
 in that it can act both as a fundamental field, the
gauge boson of QED, and as an extended object with structure. The structure
function of the photon differs from that of the proton and other hadrons,
because the photon has a point-like coupling to quarks, calculable in
perturbative QCD, as well as a non-perturbative hadron-like component, described
by the vector meson dominance model (VDM) as a superposition of the light 
neutral vector mesons $\rho^0$, $\omega^0$ and $\phi$.

The classical way to study the  structure of the photon~\cite{Nisius} 
at $\mathrm{e^+e^-}$ colliders 
is through the deep inelastic scattering (DIS) 
of electrons (or positrons) on the 
quasi-real photons emitted by the other beam.
The structure functions $F^\gamma_2$ and $F_{\mathrm{L}}^{\gamma}$ 
can be extracted from the measured differential
cross-section using the following formula:
$$
\frac{\mbox{d}^2\sigma_\mathrm{ e\gamma\rightarrow e X}}
{\mathrm{d}x\mathrm{d}Q^2} = 
  \frac{2\pi\alpha^2}{x\,Q^{4}}
  \left[\left( 1+(1-y)^2\right) 
  F_{2}^{\gamma}(x,Q^{2})
   - y^{2}
  {F_{\mathrm{L}}^{\gamma}(x,Q^{2})}\right],
$$
 where $\alpha$ is the fine structure constant, 
$Q^2$ is the photon virtuality defined as the negative four-momentum squared
of the virtual probe photon, while
 $x$ and $y$ are the
dimensionless Bjorken variables.
In the usual kinematic conditions at LEP $y$ can be
neglected, therefore the longitudinal component
$F^\gamma_L$ can not be measured.
$Q^2$ can be calculated from
the measured  angle and energy of the
scattered electron, but in contrast to electron-proton scattering,
the target photon energy is not known, therefore
$x$ can only be obtained from the invariant mass of the hadronic 
system, $W$, as
$x \approx \frac{Q^2}{Q^2+W^2},$
 neglecting the virtuality of the target photon.

\section{Photon structure at low $x$}

The most important uncertainty in the measurement of the photon structure at low
Bjorken $x$ lies in the reliable modelling of the hadronic final state, 
which in most cases is only partially contained in the detector,
leading to the need for a complicated unfolding procedure 
relying on Monte Carlo simulation.

The available Monte Carlo models have recently been critically  compared to the
combined data of several LEP experiments~\cite{LEPGG}. 
The differences between the experiments are used to estimate the systematic
errors, usually smaller  than the differences between the models, thus
the comparison can serve as a constraint when developing new models.


\FIGURE[b]{\epsfig{file=fig1.epsi,width=0.6\textwidth,height=0.5\textwidth}%
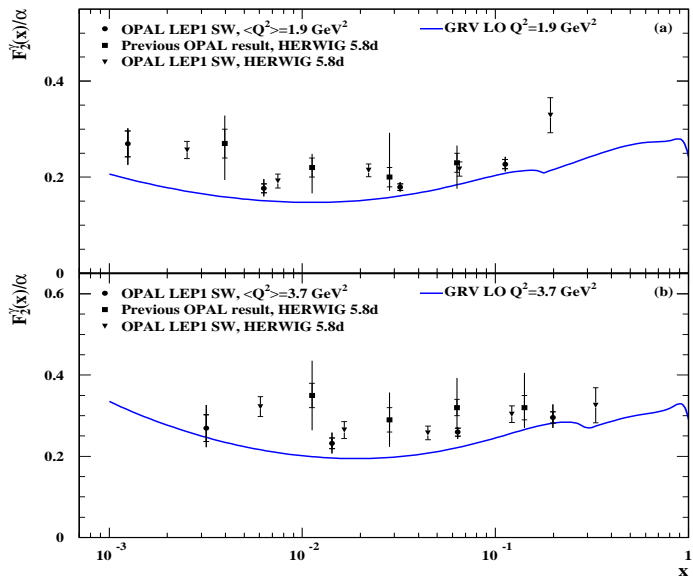
        \caption{Photon structure function measured at the lowest $x$ attainable
	at LEP. The inner error bars show the statistical error, while the outer
	error bars, where shown, correspond to the total error.}%
        \label{low-x}}

OPAL recently published a 
measurement~\cite{OPAL-low-x}  of $F_2^\gamma$  concentrating on its 
behaviour at low $x$ in various regions of 
$\langle Q^2\rangle$ from 1.9 to 17.8
GeV$^2$. 
This analysis improves the reconstruction of the hadronic final state
by incorporating kinematic information from the scattered electron, measured
more precisely than the hadrons, and uses a special treatment  of the  energy
in the forward calorimeters, to make the detector response more uniform. 
In addition a two-variable unfolding 
in $x_\mathrm{cor}$, the measured value of the Bjorken variable $x$
using the above methods,
and  $E_\mathrm{T}^\mathrm{out}/E_\mathrm{tot}$, 
the transverse energy out of the plane of the scattered electron scaled by the
total observed energy, is used to further reduce the model
dependence of the results, leading to  much reduced total systematic errors, 
as can be seen in Figure~\ref{low-x} for the two lowest $Q^2$ regions.

The data are consistent with, but do not prove the presence of a rise in
$F_2^\gamma$ at low $x$ predicted by QCD and observed for the proton at HERA.
In general, the shape of the GRV and other leading order parametrisations 
is consistent with the data, but the predictions are too low for low
$\langle Q^2\rangle$ values.

\section{Photon structure at high $Q^2$}

At high $Q^2$ the point-like component of $F_2^\gamma$ is expected to dominate.
In this region the hadronic system has more transverse momentum and therefore
it is better contained in the detector, leading to much better correlation
between its true and measured invariant mass. This fact justifies the use of
the traditional one-dimensional unfolding  in the  $x_\mathrm{vis}$  variable,
the value of $x$ calculated from the  part of the  hadronic  final state that
is visible in the detector.  This approach was used in the new OPAL
measurement~\cite{OPAL-low-x} using the full LEP2 data set
shown in Figure~\ref{highq2}, an extension of the
measurement discussed above to  the $Q^2$ range from 7.1 to 2323 GeV$^2$ by
detecting the scattered electron in  the electromagnetic endcap detectors.
\FIGURE{\epsfig{file=fig2.epsi,width=0.6\textwidth}%
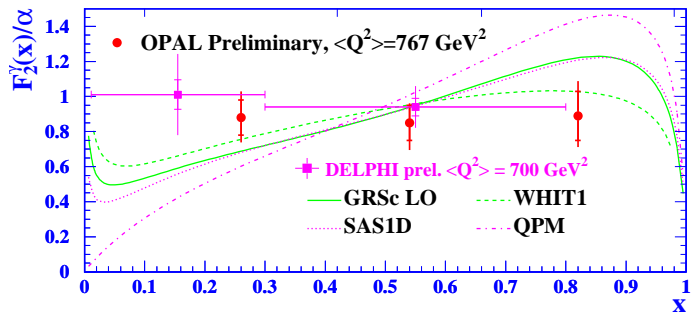
        \caption{Measured values of the hadronic photon structure function 
	$F^\gamma_2$ at the highest $\langle Q^2\rangle$ measured so far, 
	compared to various parametrisations.}%
        \label{highq2}}

A similar measurement~\cite{DELPHI} performed by DELPHI is also shown in 
Figure~\ref{highq2}. This analysis uses a new approach with a multi-variable
fit to the  observed distributions to adjust the individual components of the
hadronic structure function.  This method results in larger uncertainties, but
reduced model dependence.

\section{$Q^2$ evolution of the photon structure function}

Perturbative QCD can not predict the absolute normalisation of $F^\gamma_2$, 
which has to be extracted from data, but its evolution with $Q^2$ is predicted
to be logarithmic. Previous measurements of this evolution have been improved
significantly by DELPHI and OPAL
in~\cite{DELPHI}  and~\cite{OPAL-high-q2}  both in terms of the
precision and in terms of the $Q^2$ range covered, as shown in
Figure~\ref{evolution} for medium values of $x$.
The measured slope is  noticeably higher than the existing 
leading order QCD parametrisations.
\FIGURE{\epsfig{file=fig3.epsi,width=0.6\textwidth}%
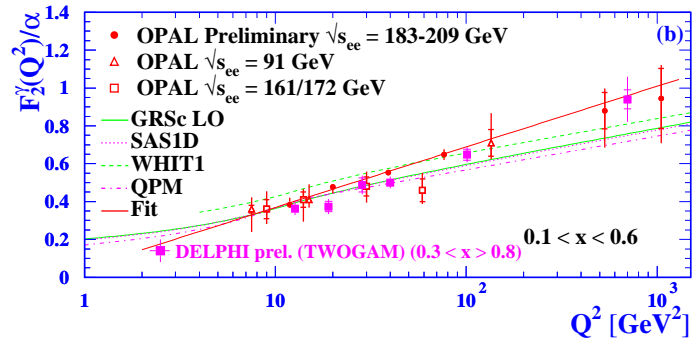
        \caption{The evolution of %
	$F^\gamma_2$ with $Q^2$ at medium values of $x$, %
	compared to various parametrisations.}%
        \label{evolution}}

Instead of one wide range of $x$,  the $Q^2$ evolution of $F^\gamma_2$ can
also  be measured in separate regions of $x$  so that the scaling violations 
in each region can be examined. 
In contrast to the proton, the photon structure
function shows  positive scaling violations at all values of $x$, as expected
from QCD due to the point-like coupling of the photon to quarks.

\section{Charm structure function of the photon}

An interesting question is the flavour composition of the photon structure
function. 
Charm events can be recognised by reconstructing the $\mathrm{D^\star
\to D^0\pi}$ decay, where the small mass difference between the D$^\star$ and
the D$^0$ 
ensures that the phase
space for random combinations faking this decay is small. Applying this well
established charm identification method to deep inelastic e$\gamma$ scattering
one can measure the charm structure function of the photon separately.

\FIGURE{\epsfig{file=fig4.epsi,width=0.6\textwidth,height=0.5\textwidth}%
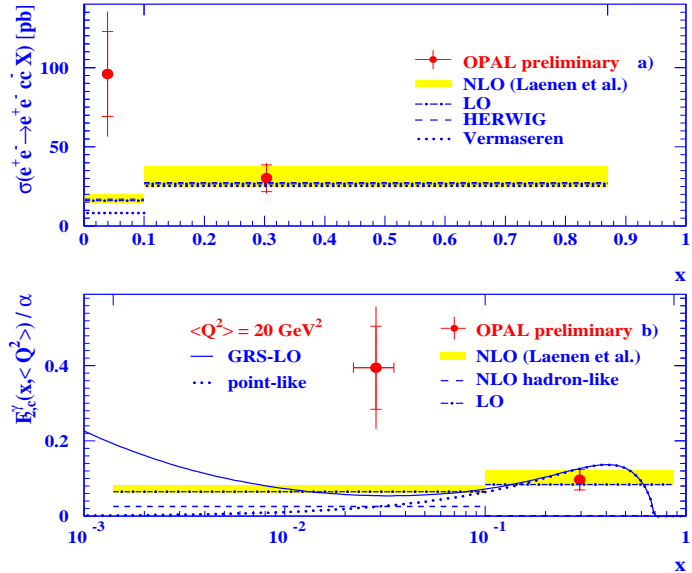
        \caption{The total cross section of charm production 
	in e$\gamma$ DIS and the charm structure function of the photon.}%
        \label{f2c}}
Figure~\ref{f2c} shows the recent update~\cite{OPAL-charm} of  the first such
measurement performed by OPAL, using improved Monte Carlo models and all the
LEP2 data  to obtain better precision.  The data are divided in two bins of
$x$; the  point-like component of the photon is expected to
dominate for $x>0.1$, while the hadron-like component is more significant for
$x<0.1$, according to the NLO calculation of~\cite{charm-calc}.

While the OPAL data are well described by both the NLO calculation and the
leading order Monte Carlo models for $x$ above 0.1, the predictions are  much
lower than the data below $0.1$, suggesting a significant hadron-like
component of $F^\gamma_\mathrm{2,c}$.

\section{Conclusions}

We have seen that the structure function of the photon has been investigated 
in a wide range of $x$ and $Q^2$ and its charm content has been studied
separately. The introduction of new experimental methods and better Monte Carlo
models coupled with the increased LEP energy and luminosity have opened up new
regions of the phase-space and lead to more precise results. Further
improvements both in the analysis techniques and in the Monte Carlo models, as
well as the combination of the data of the LEP experiments can lead to more
interesting results over the coming years.

\end{document}